\documentclass[10pt,twocolumn]{article}
\usepackage{graphicx}
\usepackage{amsmath} 
\usepackage{amssymb} 
\usepackage{fancyvrb} 
\usepackage{soul}
\usepackage{fancyhdr}
\usepackage{hyperref}
\usepackage{caption}
\makeatletter
\renewcommand{\maketitle}{\bgroup
\begin{flushleft}
  \begin{Huge}
  \textbf{\@title}\\
  \end{Huge}
  \vspace{1cm}
  \@author
\end{flushleft}\egroup
}
\makeatother
\title{Study of Di-muon Production Process in $pp$ Collision in CMS Data from Symmetry Scaling Perspective}
\author{%
    \textbf{{\large Susmita Bhaduri}}$^{1}$, \textbf{{\Large Anirban Bhaduri}}$^{2}$, \textbf{{\Large Dipak Ghosh}}$^{3}$\\
    $^{1,2,3}$Deepa Ghosh Research Foundation, Kolkata-700031,India \\
    \underline{$^{1}$susmita.sbhaduri@dgfoundation.in}\\
    \underline{$^{2}$bhaduri.anirban@dgfoundation.in}\\
    \underline{$^{3}$deegee111@gmail.com}
}

\begin{document}
\twocolumn[
  \begin{@twocolumnfalse}
    \maketitle
  \end{@twocolumnfalse}
  ]
\noindent

\date{\today}

\begin{abstract}

The analysis of relativistic $AA$ collisions is known to give rise to observables which would bear the signature of Quark-Gluon-Plasma(QGP) in the relativistic heavy ion collisions. An extensive knowledge of $pp$ collisions is required both as input to comprehensive theoretical models of strong interactions and also as a baseline to decipher the $AA$ collisions at relativistic and ultrarelativistic energies, which has been of great interest in the area of theoretical and experimental physics. The multiplicity distribution of particles generated in $pp$ collisions and the multiplicity dependence of various global event features serve as rudimentary observables which reflect the features of the underlying dynamics of the process of particle production. Moreover, the the outcomes of the assessment of $pp$ and $pA$ systems, should be used as a reference to validate the understanding of the processes which are expected to contribute to the dynamics of the process of di-muon production~\cite{Uras2016}. 
Recent availability of di-muon data has triggered spur of interests in revisiting strong interaction process, the study of which in detail is extremely important for enhancement of our understanding on not only the theory of strong interaction but also possible physics scenarios beyond the standard model. A good number of papers have come up where background of production process of di-muon in $pp$ collision has been discussed and analyzed particularly for production of dimuon from $\gamma\gamma$ interaction. Several other possibilities are also taken care of. However, future data of $pp$ collision will provide opportunities to validate different approaches~\cite{Goncalves2018}. Apart from conventional approaches to study dynamics of particle production in high energy collision the present authors proposed a new approach with successful application in context of symmetry scaling in $AA$ collision data from ALICE Collaboration~\cite{alice} in~\cite{Bhaduri20184}, $pp$ collisions at $8$TeV from CMS collaboration~\cite{cms2017} in~\cite{Bhaduri20191} and also in other numerous works with different collision data.
This different approach essentially analyses fluctuation pattern from the perspective of symmetry scaling or degree of self-similarity involved in the process. 
The proposed methods of analysis using single variable of pseudorapidity values of di-muon data taken out from the primary dataset of RunA(2011) and RunB(2012) of the $pp$ collision at $7$TeV and $8$TeV respectively from CMS collaboration~\cite{cms2017}, reveal that pseudorapidity spaces corresponding to different ranges of pseudorapidity-values are highly scale-free and possess fractal characteristics. They also reveal that the scaling pattern changes from one rapidity range to another and also from one range of energy to another. Further, it has been found that all the pseudorapidity spaces are highly cross-correlated with their corresponding azimuthal-spaces and their degree of cross-correlation varies from one range of rapidity to other and also from one range of energy to another.
\end{abstract}

\textbf{Keywords:} Symmetry scaling, Multifractal analysis, Multifractal cross-correlation, Di-muon production.\\                             
\textbf{PCAS Nos.:} \textit{10}

\section{Introduction}
\label{intro}
In the recent past, fluctuation and correlation have been analyzed widely using novel methods of studying non-statistical fluctuation which resulted in the better understanding of the dynamics of the pionisation process. The methods including the process of intermittency were introduced by Bialas and Peschanski~\cite{bialash986} and they observed association between intermittency indices and anomalous fractal dimension~\cite{bia1ash1988,dewolf1996}. Then the parameters of Gq moment and Tq moment~\cite{hwa90,paladin1987,Grass1984,hal1986,taka1994} were introduced which were deduced from various methods based on fractal concepts. Then distinctive approaches of \textit{Detrended Fluctuation Analysis(DFA) and multifractal-DFA(MF-DFA)}~\cite{Peng1994, kant2002} were applied extensively for analysing non-stationary, nonlinear properties of data series to investigate the long-range correlations inherent in the process of particle production~\cite{Albajar1992,Suleymanov2003,YXZhang2007,Ferreiro2012}. In various contemporary works, self-similarity has been analysed in the areas of particle physics which includes - the production process of Jet and Top-quark in the experiments of Tevatron and LHC~\cite{Tokarev2015}, the procedure of strangeness production in $pp$ collisions at the RHIC~\cite{Tokarev2016} experiments, the phenomenon of proton spin and asymmetry inherent in jet production process~\cite{Tokarev20151} and to decipher the collective phenomena~\cite{Baldina2017} and the process of establishment of the notion of self-similar symmetry of dark energy~\cite{TomohideSonoda2018}.
Study of long range cross-correlation between two non-stationary signals \textit{Detrended Cross Correlation Analysis(DXA)} had been presented by Podobnik et al.~\cite{Podobnik2008}.
Zhao et al.~\cite{Wang2013} introduced \textit{Multifractal Detrended Cross-Correlation Analysis(MF-DXA)} by combining \textit{MF-DFA} and \textit{DXA} methods to examine higher degree of multifractal parameters of two cross-correlated series.
MF-DXA method has been applied with substantially higher degree of accuracy in the analysis of the unrevealed cross-correlation in the various fields of physics, physiology finance and power markets~\cite{Podobnik2008,Wang2013} and also in the fields of particle physics~\cite{Bhaduri20171}.

We have performed the scaling analysis of the pseudorapidity space taken out from Pb-Pb VSD masterclass data at $2.76 TeV$ per nucleon pair from ALICE Collaboration~\cite{alice} using both the method of complex network based Visibility Graph and multifractal-DFA(MF-DFA)~\cite{Peng1994, kant2002}, to study the prospective phase transition and the signature of QGP ~\cite{Bhaduri2630203,Bhaduri20184}. We also studied multiplicity fluctuation process in nucleus-nucleus and hadron-nucleus interactions by applying complex network and chaos based Visibility Graph methodology in quite a few recent works~\cite{Bhaduri20167,Bhaduri20171,Bhaduri20183,Bhaduri20163,Bhaduri20165,Bhaduri20166,Bhaduri20172,Bhaduri20181,Bhaduri20191}. These techniques have also been successfully applied to identify phase transitions in temperature driven magnetization properties~\cite{Zhao2016} and also in temperature-driven phase transition from liquid to vapour state~\cite{Zebende2004}.
In a recent study~\cite{Chatrchyan2013} different combinations of topological and kinematic input variables from the data of RunA(2011) of the $pp$ collision at $7$TeV at CMS detector have been used, from which several ANNs(Artificial Neural Networks) have been constructed and then after comparison the optimally configured ANN has been selected .

The outcomes of the assessment of $pp$ and $pA$ systems, should be used as a reference to validate the understanding of the processes which are expected to contribute to the dynamics of the process of di-muon production~\cite{Uras2016}. Moreover, apart from the analysis of $AA$ collisions, an extensive knowledge of $pp$ collisions is required both as an input to comprehensive theoretical models of strong interactions and also as a baseline to decipher the $AA$ collisions at relativistic and ultrarelativistic energies. This has been of great interest in the area of theoretical and experimental physics.
The process of soft particle generation from ultrarelativistic $pp$ collisions is affected by the flavor distribution among the proton, quark hadronization and baryon number transport. In the process of $AA$ collisions, the magnitude of the spectrum of transverse momentum of charged particles in $pp$ collisions serves as an important reference. A $pp$ reference spectrum is required for $AA$ collisions to probe for the effects of probable initial states in the collision. The multiplicity distribution of particles generated in $pp$ collisions and the multiplicity dependence of various global event features serve as rudimentary observables which reflect the features of the  underlying dynamics of the process of particle production.


After few successful endeavors in the field of analyzing the pionisation process in high energy interaction using chaos-based procedures and also motivated by the different attempts reported to investigate the dynamics of the generation process of di-lepton pairs in the work~\cite{Werner2018}, we have attempted to analyze the di-muon production process in hadron-hadron interactions.
We have proposed to implement the chaos based methods of DFA, MF-DFA, MF-DXA to analyze the energy and rapidity dependence of di-muon production process by utilizing a single variable of pseudorapidity values of di-muon data taken out from the primary dataset of RunA(2011) and RunB(2012) of the $pp$ collision at $7$ TeV and $8$ TeV respectively from CMS collaboration~\cite{cms2017}. The rapidity and energy dependence of the process are examined by means of fundamental scaling parameter signifying the degree of symmetry scaling or scale-freeness in the di-muon production process, extracted by the proposed method.
All these methods reveal that pseuodorapidity spaces corresponding to different range of pseudorapidity - values are highly scale-free and possess fractal characteristics. They also reveal that how the scaling pattern changes from one rapidity range to another and also from one range of energy to another.

The rest of the paper is structured as per the followings. Section~\ref{ana} describes the methods of analysis. Section~\ref{mfdfa} presents the algorithm of DFA, MF-DFA and Section~\ref{mfdxa} the method of MF-DXA in detail and the importance of the parameters - the width of multifractal spectrum and the cross-correlation exponent. Section~\ref{data} describes the data in detail. Section~\ref{method} describes the details of our study and the deductions from the test-results. Section~\ref{con} details the physical importance of the proposed parameters and their relevance with regards to the dynamics of the di-muon production process and finally concludes the paper.

\section{Method of analysis}
\label{ana}
We have elaborated Multifractal-Detrended Fluctuation Analysis(MF-DFA) method~\cite{Peng1994, Kantelhardt2001, kant2002} to calculate the Hurst exponent and the width of the multifractal spectrum. Then we have extracted these parameters for analyzing the fluctuation of data series extracted from the experimental data as elaborated in the Section~\ref{data}.
\subsection{MF-DFA method}
\label{mfdfa}
\begin{enumerate}
\item Here we denote the experimental data series as $x(i)$ for $i = 1,2,\ldots,N$, where $N=$ number of points. The average of this series is computed as $\bar{x} = \frac{1}{N}\sum_{i=1}^{N} x(i)$. Then the collective deviation series for $x(i)$ is calculated as per the equation[~\ref{eqn1}].
\begin{eqnarray}
\label{eqn1}
X(i) \equiv \sum_{k=1}^{i} [x(k)-\bar{x}], i = 1,2,\ldots,N  
\end{eqnarray}
This deduction of the average($\bar{x}$) from the input data series, is a conventional method of eliminating noise from the input data series. The result of this subtraction would be removed by the detrending process in the fourth step.

\item $X(i)$ is then divided into $N_s$ non-overlapping segments, with $N_s \equiv int(N/s)$ and $s$ as the length of the segment. In this experiment $s$ ranges from $16$(minimum) to $1024$(maximum) value in log-scale.

\item For each $s$, a particular segment is denoted by $v$($v = 1,2,\ldots,N_s$). Least-square fitting is performed for each segment to derive the local trend for that specific segment~\cite{Peng1994}.
$x_v(i)$ denotes the least-square fitted polynomial for the segment $v$ in series $X(i)$. $x_v(i)$ is computed according to the equation $x_v(i) = \sum_{k=0}^{m} {C_{k}}{(i)^{m-k}}$, with $C_{k}$ as the $k$th coefficients of the fitted polynomial of degree $m$. Different kinds of fitting - linear, quadratic, cubic or higher $m$-order polynomial, may be used~\cite{Kantelhardt2001, kant2002}. In this experiment linear least-square fitting is applied with $m=1$.

\item Now to detrend the data series, the least-square fitted polynomial is subtracted from the data series. There is existence of slow-varying trends in natural data series. So to extract the scale invariant structure of the dissimilarity around the trend, detrending is necessary.
For each value of $s$ and segment $v \in 1,2,\ldots,N_s$, detrending is executed by deducting the least-square fit $x_v(i)$ from the specific portion of the data series $X(i)$, for the segment $v$ to calculate the variance which is denoted by $F^2(s,v)$ computed as per the equation[~\ref{eqn2}].
\begin{eqnarray}
\label{eqn2}
F^2(s,v) \equiv \frac{1}{s}\sum_{i=1}^{s} \{X[(v-1)s+i]-x_v(i)\}^2
\end{eqnarray}
with $s \in 16,32,\ldots,1024$ and $v \in 1,2,\ldots,N_s$.

\item Next, the $q$th-order function of fluctuation, denoted by $F_q(s)$, is computed by averaging the values of $F^2(s,v)$ over all the segments($v$) produced for each $s \in 16,32,\ldots,1024$ and for a specific $q$, as per the equation[~\ref{eqn3}].
\begin{eqnarray}
\label{eqn3}
F_q(s) \equiv \left\{\frac{1}{N_s}\sum_{v=1}^{N_s} [F^2(s,v)]^{\frac{q}{2}}\right\}^{\frac{1}{q}}
\end{eqnarray}
Here $q \neq 0$ as in that case $\frac{1}{q}$ would blow up. In this experiment $q$ varies from $(-5)$ to $(+5)$. For $q = 2$, computation of $F_q(s)$ would sum up to conventional method of Detrended Fluctuation Analysis(DFA)~\cite{Peng1994}.

\item \label{seqn4}The above steps are repeated for various values of $s \in 16,32,\ldots,1024$ and it is observed that for a particular $q$, $F_q(s)$ rises in value with increasing $s$. If the data series is long range power correlated, then $F_q(s)$ vs $s$ for a specific $q$, will display power-law behavior as per the equation[~\ref{eqn4}].
\begin{eqnarray}
\label{eqn4}
F_q(s) \propto s^{h(q)}
\end{eqnarray}
If this type of scaling exists then $\log_{2} [F_q(s)]$ depends on $\log_{2} s$ in a linear fashion, where $h(q)$ is the slope which is dependent on $q$. $h(2)$ is alike to the so-called \textbf{Hurst exponent}~\cite{Kantelhardt2001}. So, $h(q)$ is defined as the generalized Hurst exponent.

\item The scaling pattern of the variance $F^2(s,v)$ is same for all segments in case of a monofractal series. In other words, the averaging of $F^2(s,v)$ would show uniform scaling behavior for various values of $q$ and hence $h(q)$ becomes independent of $q$ for monofractals. 

But, if large and small fluctuations in the series have varying scaling pattern, then $h(q)$ becomes substantially dependent on $q$. In these cases, for positive values of $q$, $h(q)$ delineates the scaling pattern of the segments with large fluctuations and for negative values of $q$, $h(q)$ describes scaling behavior of the segments with smaller fluctuations. The generalized Hurst exponent $h(q)$ for a multifractal data series is associated with the classical multifractal scaling exponent $\tau(q)$ according to the equation[~\ref{eqn5}].
\begin{eqnarray}
\tau(q) = qh(q)-1 
\label{eqn5}
\end{eqnarray}

\item As multifractal series have numerous Hurst exponents, so $\tau(q)$ depends nonlinearly upon $q$~\cite{Ashkenazy2003}. The singularity spectrum, here denoted by$f(\alpha)$, is associated with $h(q)$ as per the equation[~\ref{eqn6}].
\begin{eqnarray}
\alpha = h(q)+qh'(q), f(\alpha) = q[\alpha-h(q)]+1 
\label{eqn6}
\end{eqnarray}

Here the singularity strength is denoted by $\alpha$ and $f(\alpha)$ describes the dimension of the subset series denoted by $\alpha$. Different values of $f(\alpha)$ for different $\alpha$ results into multifractal spectrum of $f(\alpha)$ which is an arc and for this spectrum the gap between the maximum and minimum values of $\alpha$, is the \textbf{width of the multifractal spectrum} or the measurement of the multifractality of the input data series.

\item For $q=2$, if $h(q)$ or $h(2)=0.5$ then no correlation exists there in the data series. There is persistent long-range cross-correlations in the data series, which means a large value is the series is presumably to be followed by another large value in the series, if $h(2) > 0.5$. Whereas for $h(2) < 0.5$ there would be anti-persistent long-range correlations which implies a large value would probably be followed by a small value in the series and alse vice versa.

\end{enumerate}

\subsection{MF-DXA method}
\label{mfdxa}
Zhao et al.~\cite{Wang2013} have introduced \textit{MF-DXA} method based on the \textit{MF-DFA} method\cite{Kantelhardt2001,kant2002} and analyzed the cross-correlation between two non-stationary series quantitatively. 
The broad steps for the \textit{MF-DXA} method are as follows.
\begin{enumerate}
\item Let $x(i)$ and $y(i)$ are two data series for $i = 1,2,\ldots,N$, of length $N$. The mean of 
these series is calculated as $\bar{x} = \frac{1}{N}\sum_{i=1}^{N} x(i)$ and $\bar{y} = \frac{1}{N}\sum_{i=1}^{N} y(i)$ respectively.
Then accumulated deviation series for $x(i)$ and $y(i)$, are calculated as per the equation[~\ref{eqn1}]
and denoted by $X(i)$ and $Y(i)$ respectively.
Then both $X(i)$ and $Y(i)$ are divided into $N_s$ non-overlapping segments, where $N_s$  = $int(N/s)$, $s$ is the length of the segment. In our experiment $s$ varies from $16$ as minimum to $512$ as maximum value in log-scale. 

\item For each $s$, we denote a particular segment by $v$($v = 1,2,\ldots,N_s$).
Here $x_v(i)$ and $y_v(i)$ denote the least square fitted polynomials for the segment $v$ in $X(i)$ and $Y(i)$ respectively. $x_v(i)$ and $y_v(i)$ are calculated as per the equations $x_v(i) = \sum_{k=0}^{m} {C_{x_k}}{(i)^{m-k}}$ and $y_v(i) = \sum_{k=0}^{m} {C_{y_k}}{(i)^{m-k}}$, where $C_{x_k}$ and $C_{y_k}$ are the $k$th coefficients of the fit polynomials with degree $m$. For this experiment $m$ is taken as $1$~\cite{Wang2013}.

For each $s$ and segment $v$, $v = 1,2,\ldots,N_s$, detrending is done by subtracting the least-square fits $x_v(i)$ and $y_v(i)$ from the part of the data series $X(i)$ and $Y(i)$ respectively, for the segment $v$. The covariance of the these residuals, denoted by $f^2_{xy}(s,v)$ for a particular $s$ and $v$, is then calculated as per the following equation~\cite{Wang2013}.
\begin{eqnarray}
f^2_{xy}(s,v) = \frac{1}{s}\sum_{i=1}^{s} \{X[(v-1)s+i]-x_v(i)\} \times \nonumber \\ 
\{Y[(v-1)s+i]-y_v(i)\}, \nonumber
\end{eqnarray}
for each segment $v$, $v = 1,2,\ldots,N_s$.

\item Then the $q$th-order detrended covariance, denoted by $F_{xy}(q,s)$, is calculated by averaging $f^2_{xy}(s,v)$ over all the segments($v$) generated for a particular $s$ and $q$, as per the equation below~\cite{Kantelhardt2001,kant2002,Wang2013}.
\begin{eqnarray}
F_{xy}(q,s) = \left\{\frac{1}{N_s}\sum_{v=1}^{N_s} [f^2_{xy}(s,v)]^{\frac{q}{2}}\right\}^{\frac{1}{q}}, \nonumber
\end{eqnarray}
Here $q \neq 0$ because in that case $\frac{1}{q}$ would blow up. 

\item \label{fxy}The above process is repeated for different values of $s \in 16,32,\ldots,512$ and it can be seen that for a specific $q$, $F_{xy}(q,s)$ increases with increasing $s$. If the series are long range power correlated, 
the $F_{xy}(q,s)$ versus $s$ for a particular $q$, will show power-law behaviour as below~\cite{Wang2013}.
\begin{eqnarray}
F_{xy}(q,s) \propto s^{h_{xy}(q)} \nonumber
\end{eqnarray}
If this kind of scaling exists, $\log_{2} [F_{xy}(q,s)]$ would depend linearly on $\log_{2} s$, where $h_{xy}(q)$ is the slope and represents the degree of the cross-correlation between the the data series $x(i)$ and $y(i)$. 

In general, $h_{xy}(q)$ depends on $q$. $q$ ranges from negative to positive values. For $q = 2$, method is referred as so-called method of \textit{DXA}~\cite{Wang2013}.

\item As confirmed from several experiments done by Zhao et al.~\cite{Wang2013}, if $h_{xy}(q)=0.5$ then there is no cross-correlation. Further, if $h_{xy}(q) > 0.5$ then there is persistent long-range cross-correlations, where the large value of one variable is likely to be followed by a large value of another variable in the series, whereas, $h_{xy}(q) < 0.5$ then there is anti-persistent long-range cross-correlations, where a large value of one variable is most likely to be followed by a small value and vice verse, in the series.

\item $h_{xy}(q)$ for $q=2$, i.e. $h_{xy}(2)$ is the \textit{DXA} exponent. As per Podobnik et al. the cross-correlation exponent between two non-stationary series, denoted by $\gamma_i$, is calculated as per the equation $\gamma_i = 2-2\{h_{xy}(2)\}$~\cite{Podobnik2008}. For uncorrelated data series, $\gamma_i=1$, and the lower the value of $\gamma_i$, the more correlated the data series are. 
\end{enumerate}

\section{Experimental details}
The datasets for the proposed analysis are taken out from two publicly available experimental primary datasets from CMS collaboration~\cite{cms2017}. The elaboration of the data is given in the Section~\ref{data} and the through method of the experiment, described step by step, is explained in Section~\ref{method}.
\subsection{Data description}
\label{data}
The primary dataset of the $pp$ collision at $8$ TeV in AOD format from RunB of 2012~\cite{cms2017} and another dataset of $pp$ collision at $7$ TeV in the same AOD format from RunA of 2011~\cite{cms2017} of the CMS collaboration, are taken as the source datasets for this experiment. The run numbers which are slected and qualified by CMS to be processed along with the appropriate parameters for generation of the collision datasets are provided in the links - \href{https://doi.org/10.7483/OPENDATA.CMS.C00V.SE32}{\textit{\underline{link1}}} and \href{https://doi.org/10.7483/opendata.cms.3q75.7835}{\textit{\underline{link2}}} for $8$ TeV and $7$ TeV respectively. These datasets are made available for experiment. We have extracted the pseudorapidity-$\eta$ space and corresponding azimuthal-$\phi$ space for the generated dimuons from  these runs qualified by CMS from the primary datasets in the following formats - text(.txt) and .root format. In this analysis we have utilized these pseudorapidity space and the corresponding azimuthal space from the text(.txt) file. 

\subsection{Method of analysis and results}
\label{method}
\begin{enumerate}
\item \label{eta} The pseudorapidity-$\eta$ space for each of the datasets for $8$ and $7$ TeV extracted from the primary datasets of the CMS collaboration as described in the Section~\ref{data}, is divided into following $5$ ranges of $\eta$ values
\begin{enumerate}
\item $-2.5$ to $-1.5$.
\item $-1.5$ to $-0.5$.
\item $-0.5$ to $0.5$.
\item $0.5$ to $1.5$.
\item $1.5$ to $2.5$.
\end{enumerate}
For all the $5$ ranges the $\eta$ values are extracted from the full phase-space of the two source datasets and mapped into data series. The data series is plotted with the $X$-axis denoting the sequence number of $\eta$ values and the $Y$ corresponds to the $\eta$-values corresponding to the sequence number as in the $X$-axis.

For each of these data series following values are calculated.

\begin{itemize}
\item The width of the multifractal spectrum
\item Degree of Cross-correlation between the $\eta$ space and their corresponding $\phi$ space
\end{itemize}

\item For the each of the $10$ datasets ($5$ for $8$TeV and $5$ for $7$ TeV datasets) created for the $5$ ranges of pseudorapidity values, as specified in Step~\ref{eta}, the multifractal analysis is done and the width of multifractal spectrum is calculated as per the method elaborated the Section~\ref{mfdfa}. 

\begin{figure*}[h]
\centerline{
\includegraphics[width=3.5in]{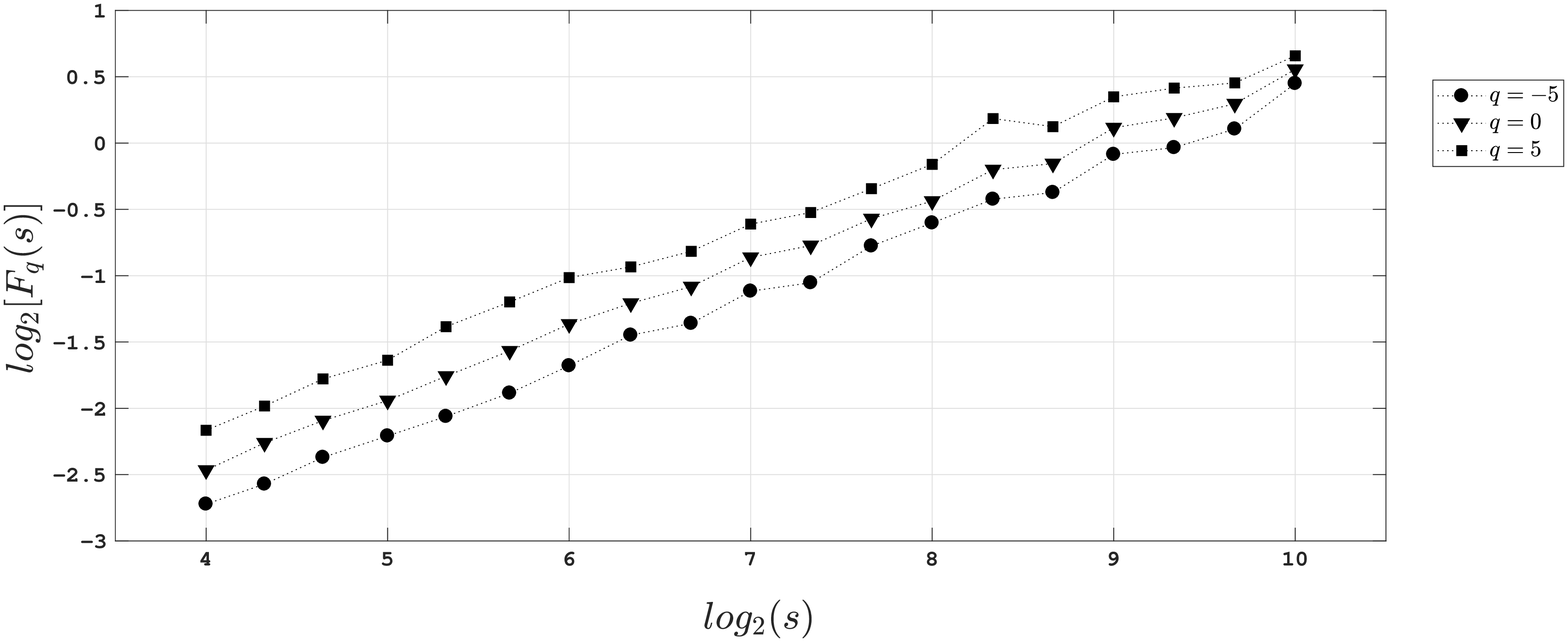}
\includegraphics[width=3.5in]{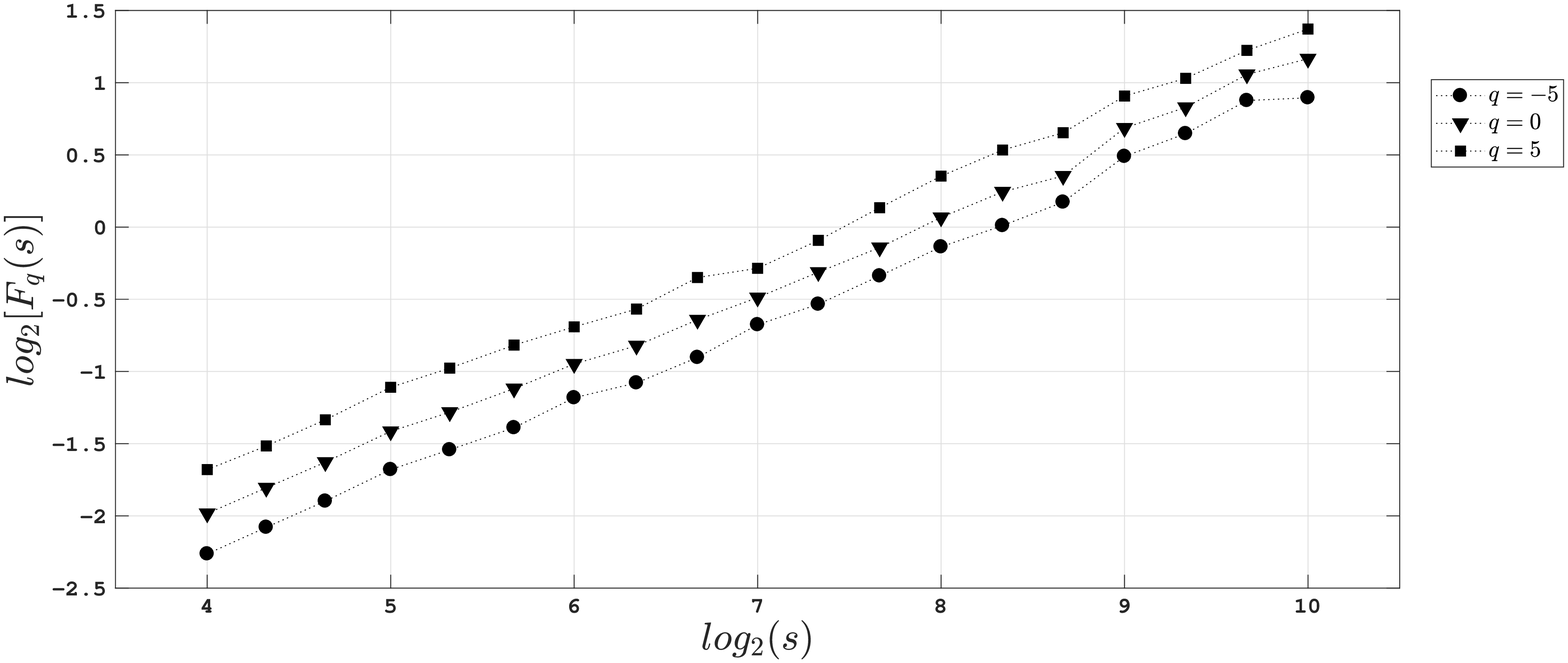}
}
\centerline{(a) \hspace*{6cm} (b)}
\caption{(a) Trend of $F_q(s)$ vs $s$ for $q = -5,0,5$, extracted for a particular range of $\eta$ for $8$ TeV dataset. (b) Trend of $F_q(s)$ vs $s$ for $q = -5,0,5$, extracted for a particular range of $\eta$ for $7$ TeV dataset.}
\label{eta78efq}
\end{figure*}

The $q^{th}$ order detrended variance $F_q(s)$ is calculated as per the Equation~\ref{eqn4} in the Step~\ref{seqn4} of the MF-DFA methodology as described in Section~\ref{mfdfa}. The Figure~\ref{eta78efq}-(a) and~\ref{eta78efq}-(b) show the $F_q(s)$ vs $s$ trend for $q = -5,0,5$, extracted for a particular range of $\eta$ values for $8$ and $7$ TeV datasets respectively. 
\begin{itemize}
\item Their linear trend confirms the power-law behavior of $F_q(s)$ versus $s$ for all the values of $q$. Similar calculation is done for all the $\eta$ ranges for both $8$ and $7$ TeV datasets and similar trend is observed.
\end{itemize}

\item For each of the $\eta$-data series corresponding to the ranges specified in Step~\ref{eta}, a randomized version of data is produced and widths of the multifractal spectrum are calculated as per the same methodology elaborated in Section~\ref{mfdfa}. The calculated values of the parameters are compared to those for the experimental data. If the source data was long range correlated, that should be eradicated by this randomization process and the data should start to be uncorrelated. Hence, as expected the widths of multifractal spectrum calculated the randomized version differ from those for the original version. In the Figure~\ref{etalpha78ac}-(a) and~\ref{etalpha78ac}-(b), the widths of the multifractal spectrum of the original datasets and their randomized versions calculated for one of the ranges $\eta$ values are shown for $8$ and $7$ TeV datasets respectively.

\begin{figure*}[h]
\centerline{
\includegraphics[width=3.5in]{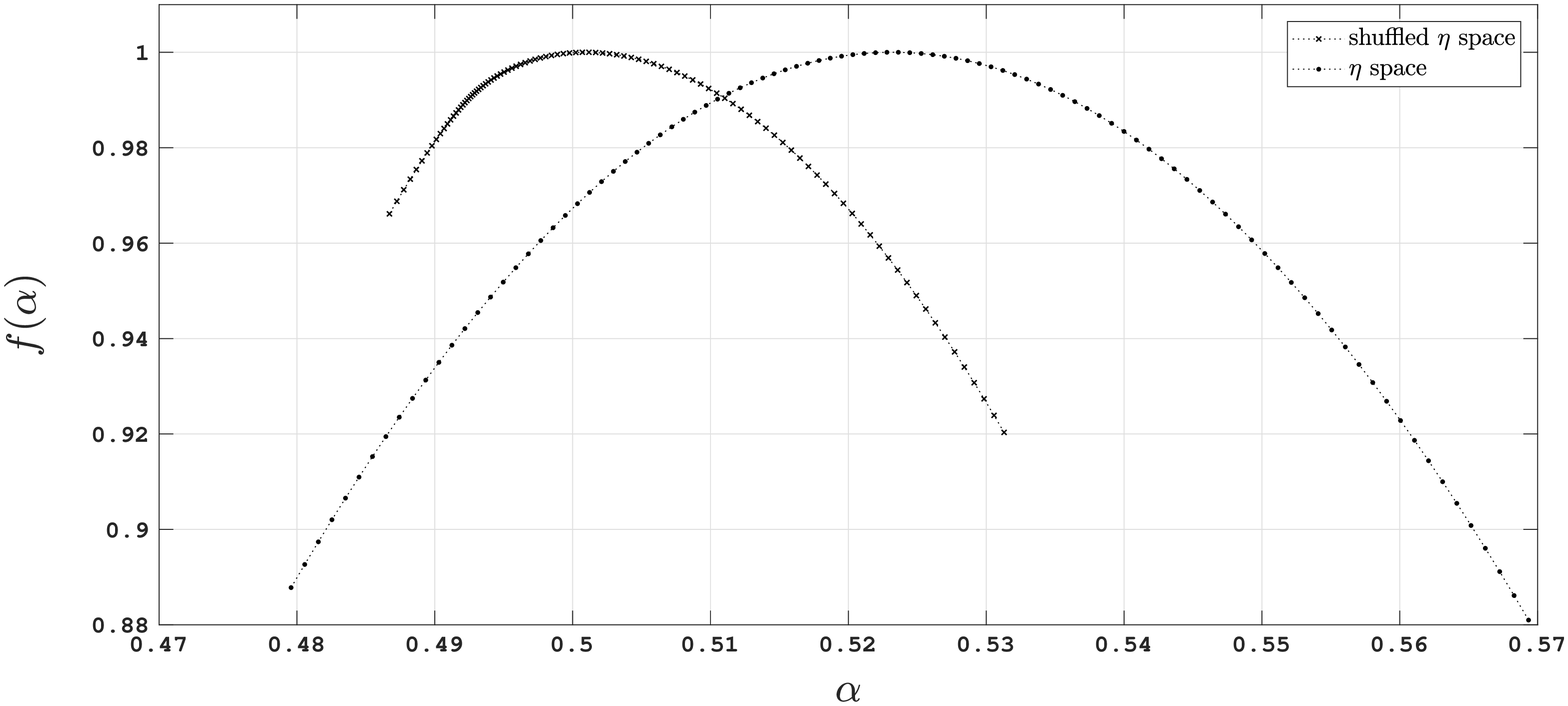}
\includegraphics[width=3.5in]{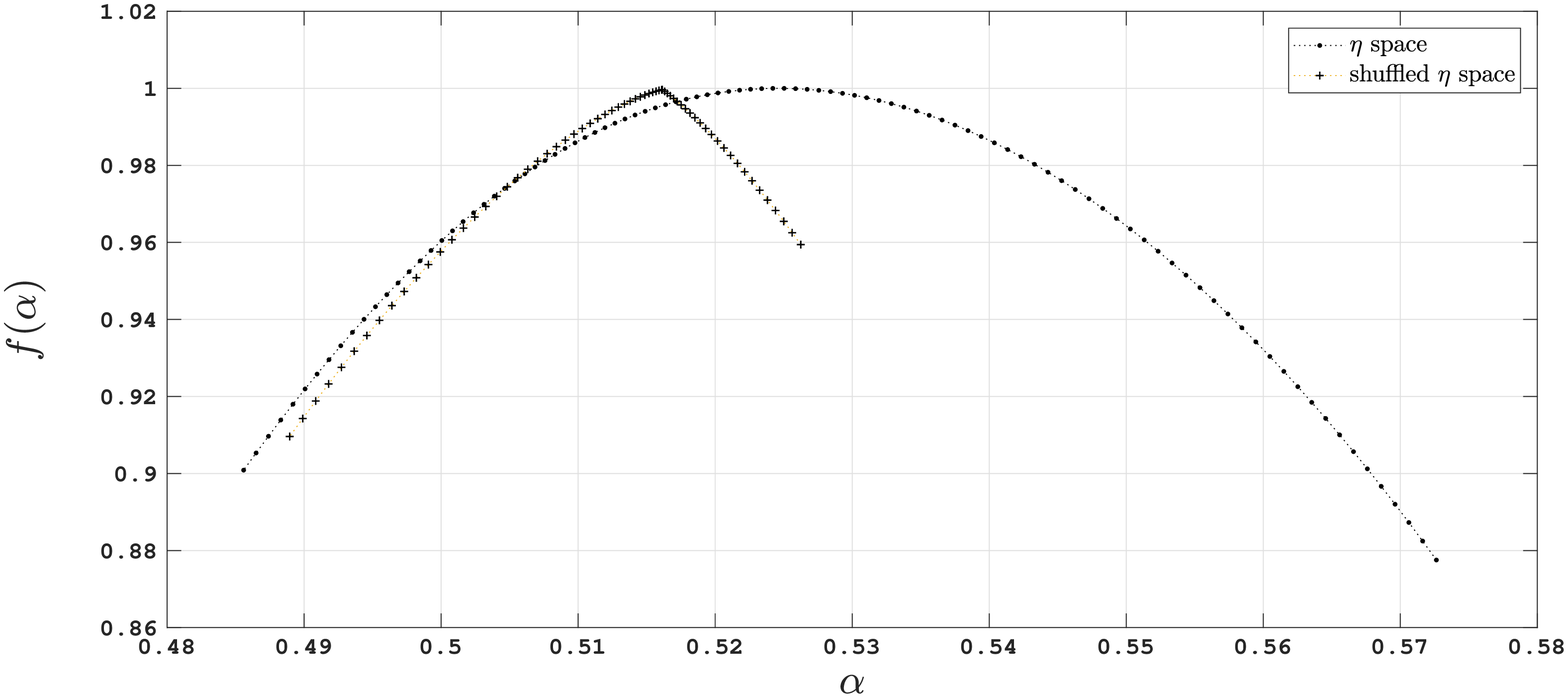}
}
\centerline{(a) \hspace*{6cm} (b)}
\caption{(a) Comparison of the trend of different values of $f(\alpha)$ versus $\alpha$ between the original and the randomized version of the $\eta$ space for a particular range of $\eta$ values for $8$ TeV dataset. (b) Comparison of the trend of different values of $f(\alpha)$ versus $\alpha$ between the original and the randomized version of the $\eta$ space for a particular range of $\eta$ values for $8$ TeV dataset.}
\label{etalpha78ac}
\end{figure*}

Similar trend is observed from the comparison of original and the randomized version of the $5$ ranges of $\eta$ values for both $8$ and $7$ TeV datasets.

Comparison of the widths of the multifractal spectrum generated for the $\eta$ spaces for all the $5$ ranges of $\eta$ values for $7$ and $8$ TeV datasets with respect to their rapidity and energy dependence is shown in the Figure~\ref{mfdfa78}. It should be noted that-
\begin{itemize}
\item The comparison of the width of the multifractal spectrum of $f(\alpha)$, denoted by the difference between the maximum and minimum values of $\alpha$, between original and the randomized version of the $\eta$ space for both the energy ranges confirm the mutifractality of the original $\eta$ spaces.
\item For the $2^{nd}$, $3^{rd}$ and $5^{th}$ range of $\eta$ values, the widths of multifractal spectrum is substantially different between the energy ranges. 
\item For both $7$ and $8$ TeV the $1^{st}$ and $4^{th}$ range of $\eta$-space displays minimum or no difference with respect to multifractality.
\item The degree of multifractality is found to be the least for $2^{nd}$ and $3^{rd}$ range for $8$ and $7$ TeV data respectively.
\end{itemize}

\begin{figure*}[h]
   \begin{minipage}{0.5\textwidth}
     \includegraphics[width=3.5in]{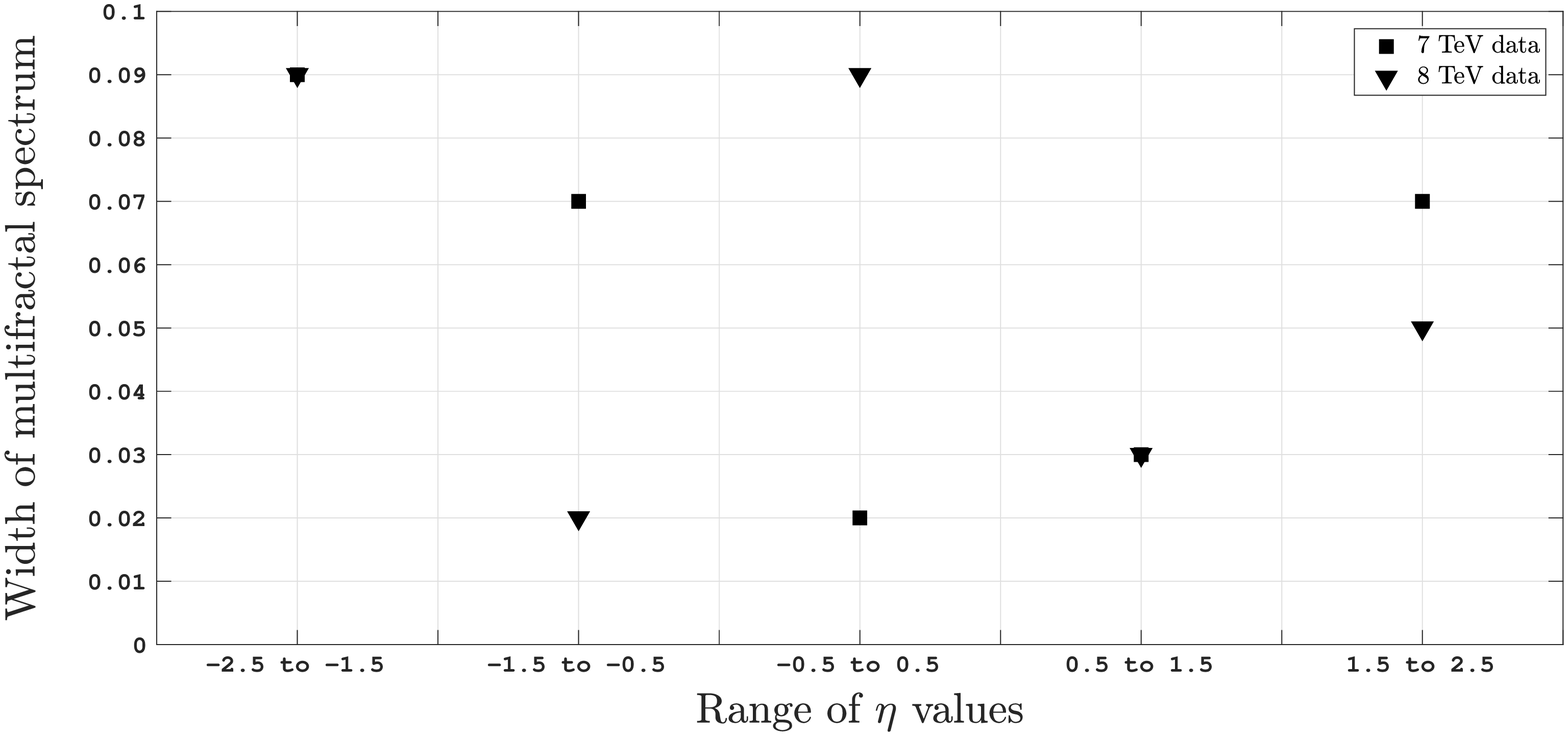}
     \caption{Comparison of the widths of the multifractal spectrum generated $\eta$ spaces for all the $5$ ranges of $		\eta$ values for $7$ and $8$ TeV datasets.}\label{mfdfa78}
   \end{minipage}
   \hspace{0.5cm}
   \begin{minipage}{0.5\textwidth}
   	\captionsetup{type=table}
     \begin{tabular}{@{}|l|c|c|c|c|@{}} \hline
      $\eta$ ranges&\multicolumn{4}{|c|}{\textbf{MFDFA Spectrum width}}\\
		\cline{2-5}
		&\multicolumn{2}{|c|}{\textbf{$8$ TeV}}&\multicolumn{2}{|c|}{\textbf{$7$ TeV}} \\
		\cline{2-5}
		&\textbf{Orig}&\textbf{Rand}&\textbf{Orig}&\textbf{Rand}\\ 
		\hline
		$-2.5$ to $-1.5$&$0.09$&$0.06$&$0.09$&$0.04$\\
		\hline
		$-1.5$ to $-0.5$&$0.02$&$0.01$&$0.07$&$0.02$\\
		\hline
		$-0.5$ to $0.5$&$0.09$&$0.04$&$0.02$&$0.04$\\
		\hline
		$0.5$ to $1.5$&$0.03$&$0.02$&$0.03$&$0.04$\\
		\hline
		$1.5$ to $2.5$&$0.05$&$0.04$&$0.07$&$0.06$\\
		\hline
     \end{tabular}
     \caption{Comparison of the widths of the multifractal spectrum generated $\eta$ spaces for all the $5$ ranges of $\eta$ values for $7$ and $8$ TeV datasets, between the original and the randomized version.}\label{gamma_mfdfa}
   \end{minipage}
\end{figure*}

Table~\ref{gamma_mfdfa} details the widths of the multifractal spectrum of the original datasets and their randomized versions for all the $10$ datasets ($5$ for $8$ and $5$ for $7$ TeV) corresponding to the $\eta$ values.

\item\label{phi} For each of the $10$ datasets ($5$ for $8$ and $5$ for $7$ TeV) of $\eta$ values extracted for the ranges specified in the Step~\ref{eta}, the corresponding azimuthal-$\phi$ space is also extracted. The $10$ corresponding $\phi$ space is sorted in the ascending order and then mapped into data series. They in turn are mapped into a two-dimensional space with their sequence along the $X$-axis and the corresponding values of $\phi$ along the $Y$ axis.

\begin{figure*}[h]
\centerline{
\includegraphics[width=3.5in]{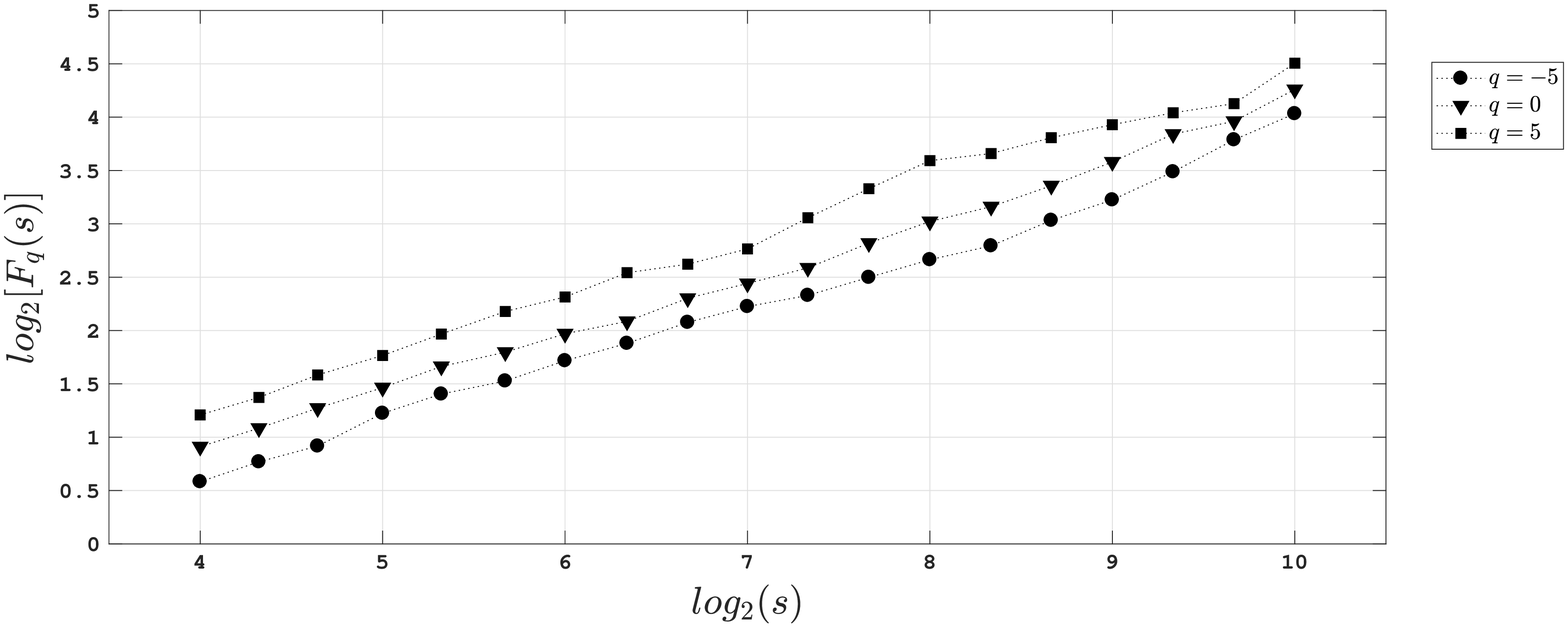}
\includegraphics[width=3.5in]{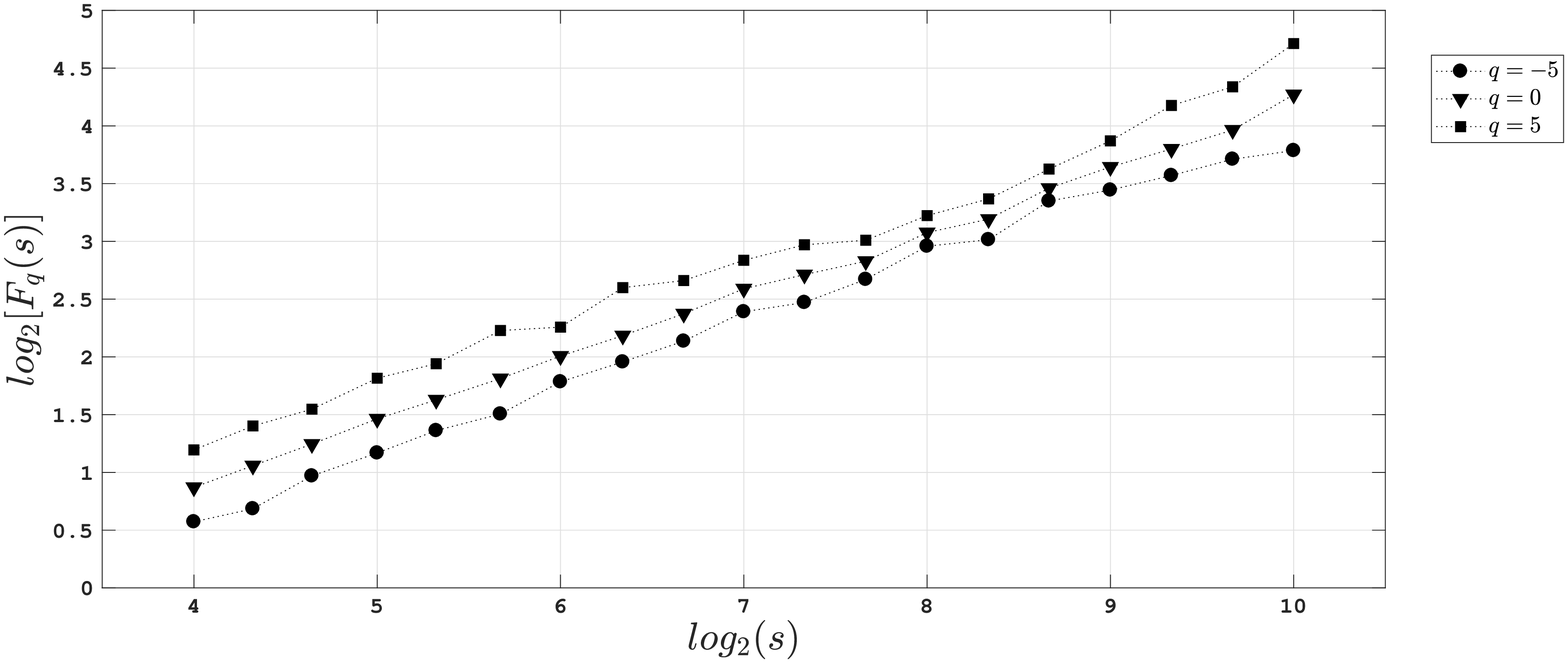}
}
\centerline{(a) \hspace*{6cm} (b)}
\caption{(a) Trend of $F_q(s)$ vs $s$ for $q = -5,0,5$, analyzed for the $\phi$ space corresponding to the $\eta$ space for $8$ TeV dataset, as shown in the Figure~\ref{eta78efq}-(a). (b) Trend of $F_q(s)$ vs $s$ for $q = -5,0,5$, analyzed for the $\phi$ space corresponding to the $\eta$ space for $7$ TeV dataset, as shown in the Figure~\ref{eta78efq}-(b).}
\label{phi78efqs_usort}
\end{figure*}

\item For the $10$ $\phi$ spaces ($5$ for $8$ and $5$ for $7$ TeV) the $q^{th}$ order detrended variance $F_q(s)$ is analyzed as per the Equation~\ref{eqn4} in the Step~\ref{seqn4} of the MF-DFA methodology as described in Section~\ref{mfdfa}. The Figure~\ref{phi78efqs_usort}-(a) and~\ref{phi78efqs_usort}-(b) show the $F_q(s)$ vs $s$ trend for $q = -5,0,5$, extracted for corresponding $\phi$ values for the same range of $\eta$ values for which same trend is shown in the Figure~\ref{eta78efq}-(a) and~\ref{eta78efq}-(b) for $8$ and $7$ TeV datasets respectively. It is to be noted that-
\begin{itemize}
\item The linear trend confirms the power-law behavior of $F_q(s)$ versus $s$ for all the values of $q$ for the $\phi$ spaces.
\item Same analysis is done for all the $\phi$ spaces corresponding to the $\eta$ ranges for both $8$ and $7$ TeV datasets and similar trend is observed.
\end{itemize}

\item Multifractal cross-correlation analysis is done as per the method described in the Section~\ref{mfdxa} between the $10$ pairs of datasets ($5$ for $8$ TeV and $5$ for $7$ TeV), one being the sorted $\phi$ values and the other being the corresponding $\eta$ values. The trend of generalized Hurst exponent($h(q)$) for different order($q$) is analyzed for all the $10$ pairs of $\eta$ and $\phi$ datasets as per the process described in Section~\ref{mfdfa}. Along with that, for the same pairs of datasets the degree of cross-correlation($h_{x,y}(q)$) for different order($q$) is analyzed as per the methodology described in Section~\ref{mfdxa}. The trend of $h(q)$ and $h_{x,y}(q)$ versus $q$ for the particular sample pair of $\eta$ and $\phi$ space for which trends of $F_q(s)$ versus $s$ are shown in Figures~\ref{eta78efq}-(a)($\eta$ space) and~\ref{phi78efqs_usort}-(a)($\phi$ space) for $8$ TeV dataset and Figures~\ref{eta78efq}-(b)($\eta$ space) and~\ref{phi78efqs_usort}-(b)($\phi$ space) for $7$ TeV dataset, are shown in the Figures~\ref{hqhxyq78}-(a) and~\ref{hqhxyq78}-(b) for $8$ and $7$ TeV datasets respectively. The values are shown in the Figures for $q=-5,-4, \ldots 5$. It should be noted that-

\begin{figure*}[h]
\centerline{
\includegraphics[width=3.5in]{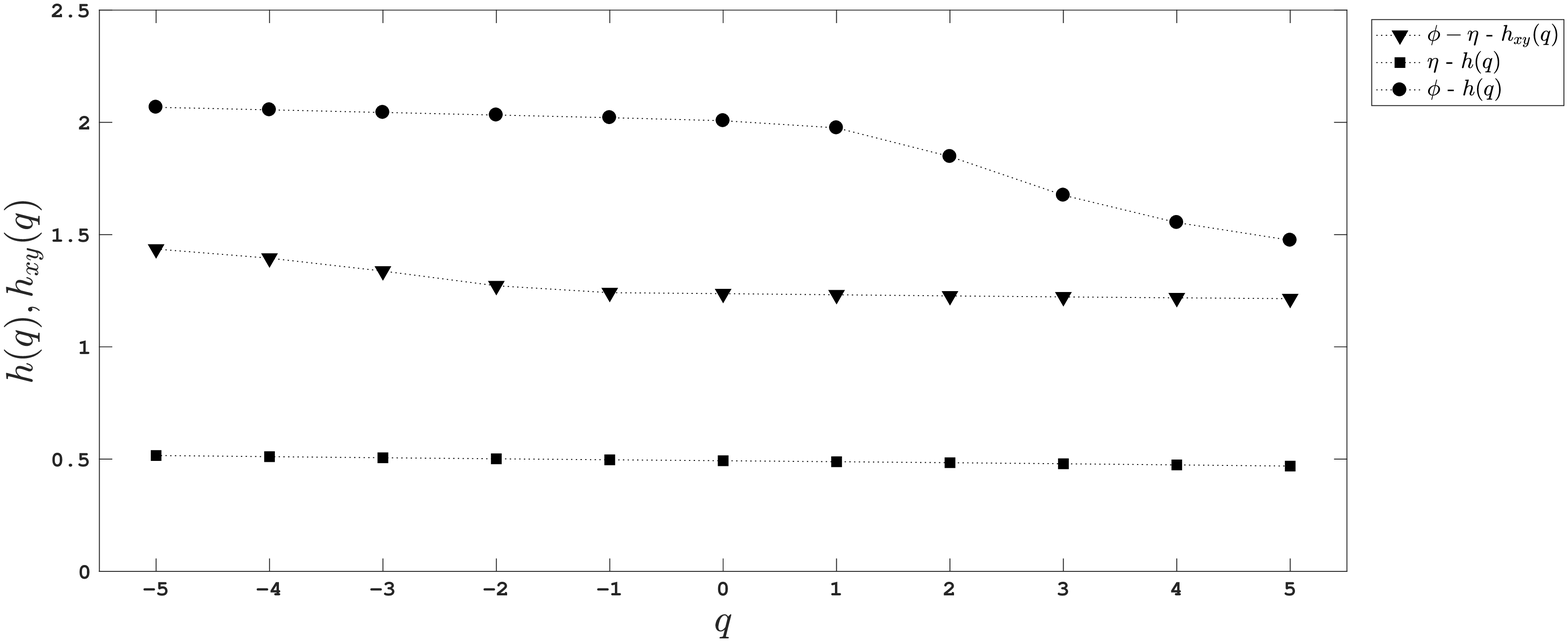}
\includegraphics[width=3.5in]{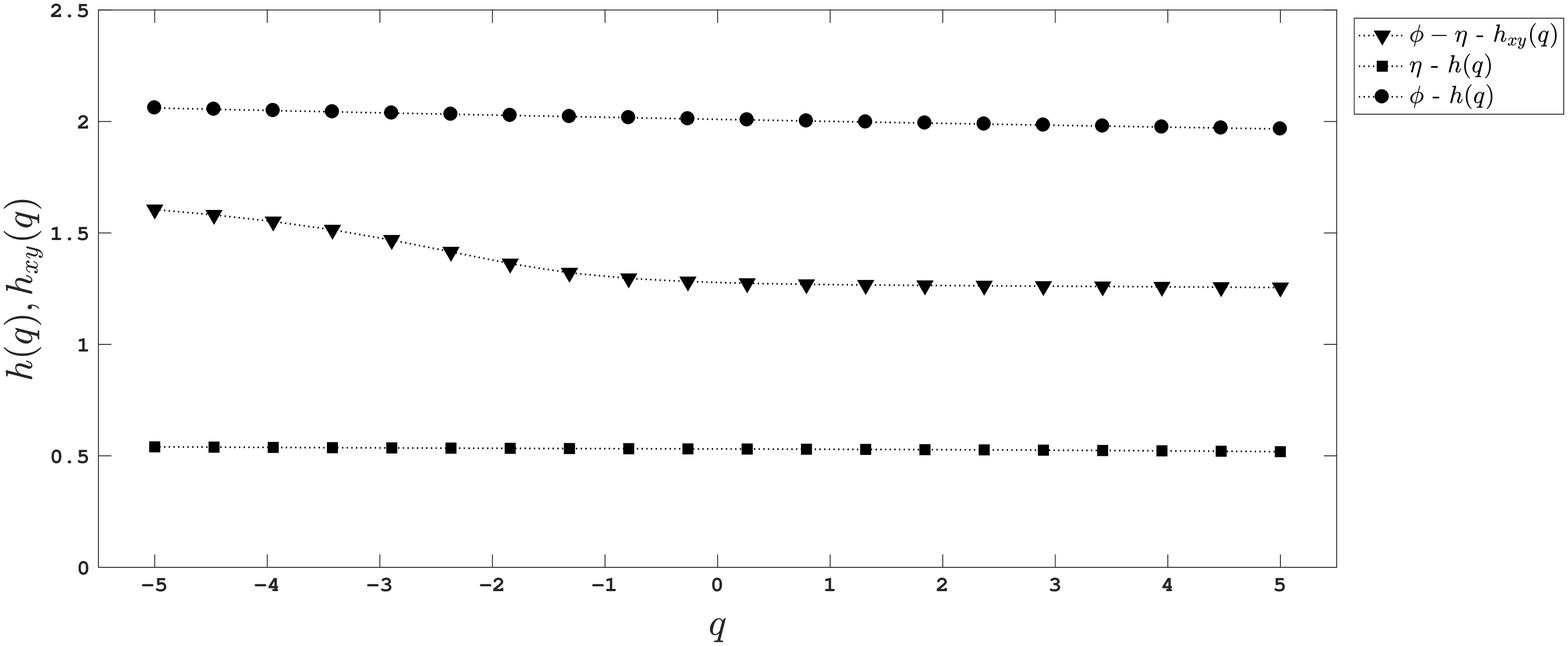}
}
\centerline{(a) \hspace*{6cm} (b)}
\caption{(a) Trend of $h(q)$ and $h_{x,y}(q)$ versus $q$ for $q=-5,-4, \ldots 5$, calculated for a particular range of $\eta$ values and their corresponding $\phi$ values for $8$ TeV dataset. (b) Trend of $h(q)$ and $h_{x,y}(q)$ versus $q$ for $q=-5,-4, \ldots 5$, calculated for a particular range of $\eta$ values and their corresponding $\phi$ values for $7$ TeV dataset.}
\label{hqhxyq78}
\end{figure*}

\begin{itemize}
\item As shown in figures~\ref{hqhxyq78}-(a) and~\ref{hqhxyq78}-(b), the trend of dependence of $h(q)$ on $q$ for individual $\eta$ and $\phi$ spaces confirm their multifractality and the same for $h_{x,y}(q)$ on $q$ for the same pair of $\eta$ and $\phi$ spaces confirm their cross-correlation for both $8$ and $7$ TeV datasets.
\item For $q=2$ both $h(q)$ and $h_{x,y}(q)$ are $>0.5$ and for $\phi$ space $h(q)$ is much higher than that for the corresponding $\eta$ space.
\item Also, $h_{x,y}(q)$ is much higher than $0.5$ for the pair of datasets for $q=2$. This suggests the presence of long-range correlation and persistence in both the spaces.
\item Moreover, there is a drop in the value of $h_{x,y}(q)$ around $q=-1$.
\item Similar analysis has been done for all the $10$ pairs of datasets and similar trend is observed for all of them.
\end{itemize}

\item Figures~\ref{alpha78e}-(a) and~\ref{alpha78e}-(b) show the comparison of the trend of different values of $f(\alpha)$ versus $\alpha$ for the same $\eta$, $\phi$ spaces and same trend calculated for their cross-correlation, for $8$ and $7$ TeV datasets respectively.

\begin{figure*}[h]
\centerline{
\includegraphics[width=3.5in]{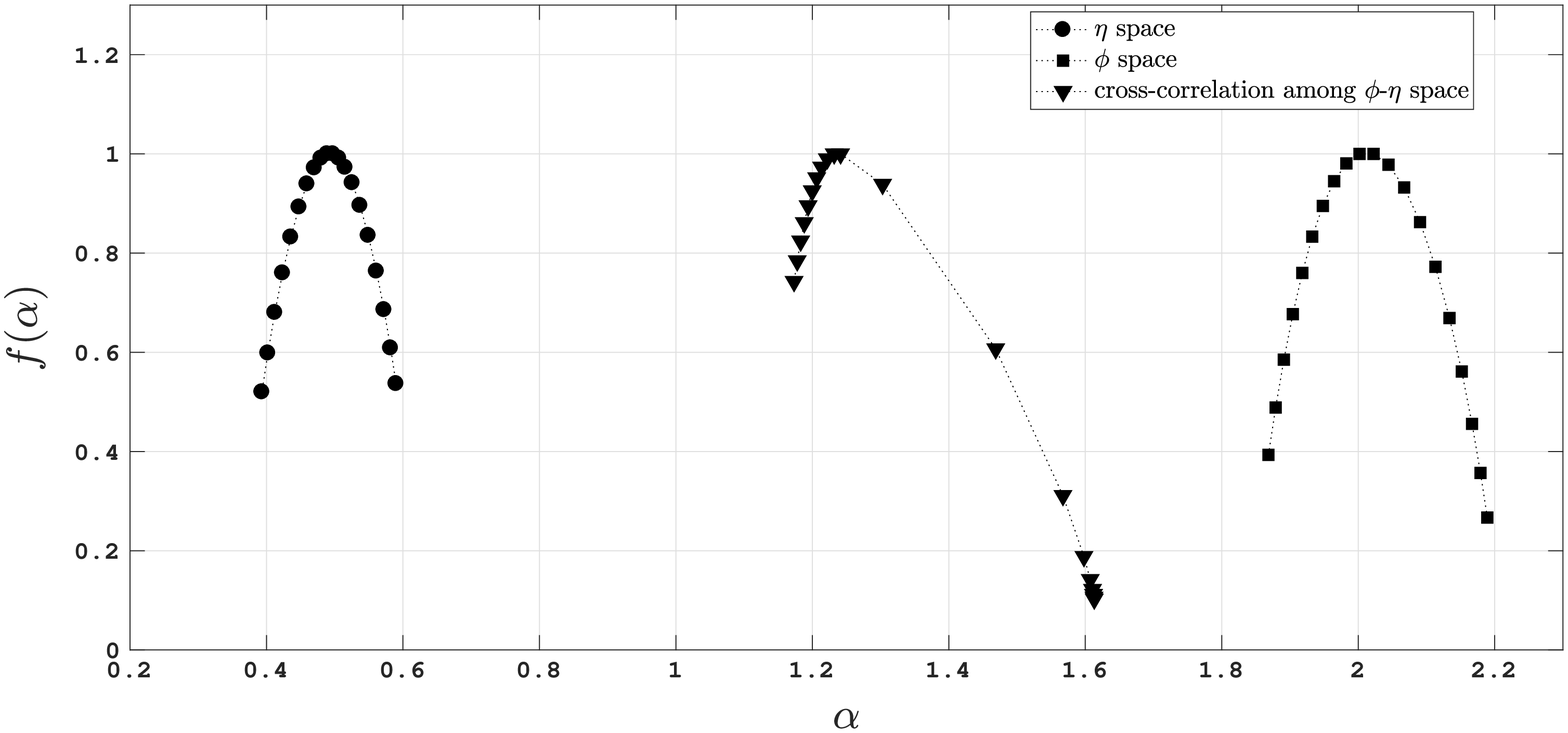}
\includegraphics[width=3.5in]{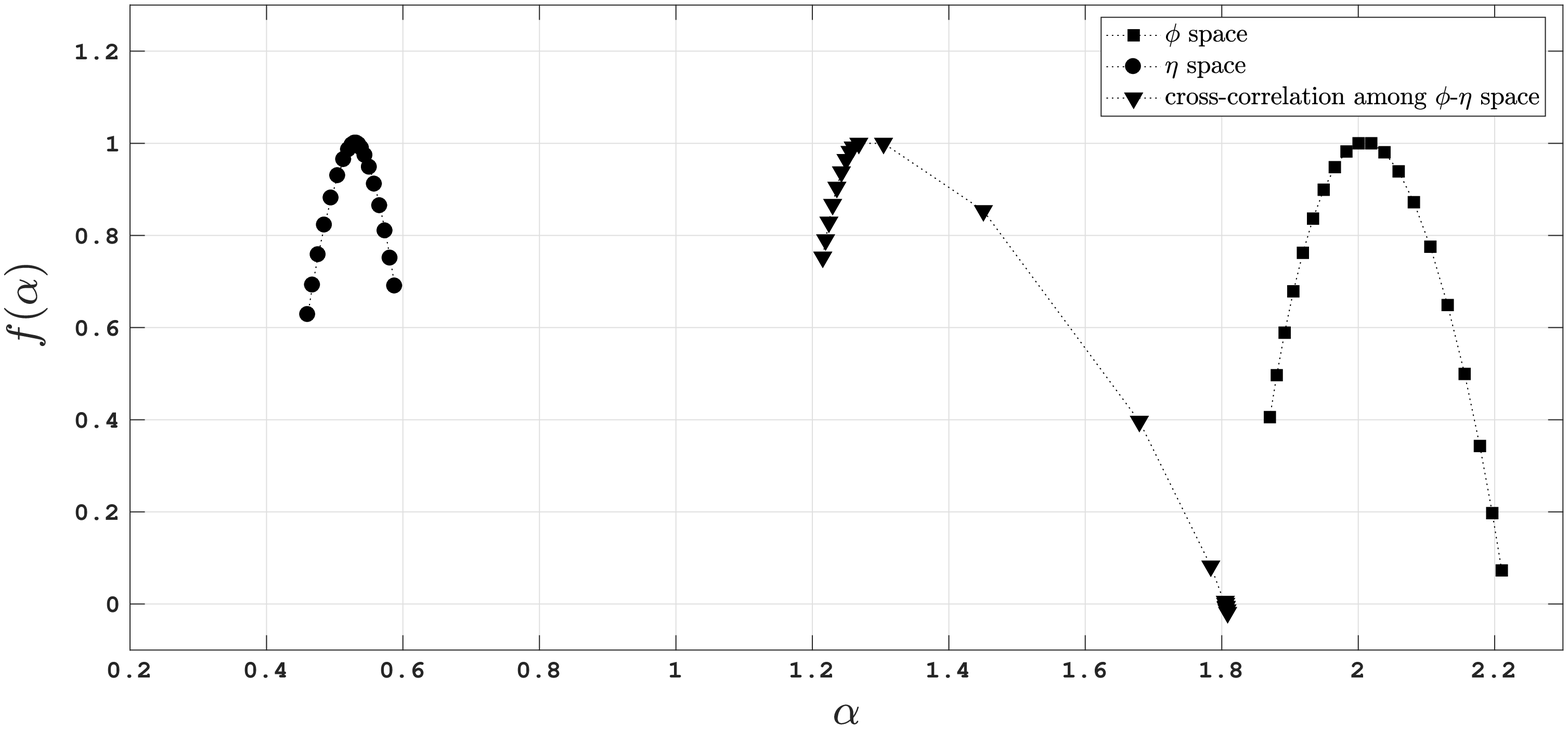}
}
\centerline{(a) \hspace*{6cm} (b)}
\caption{(a) Comparison of the trend of different values of $f(\alpha)$ versus $\alpha$ between among the same $\eta$, $\phi$ spaces and same trend calculated for their cross-correlation, for the $8$ TeV dataset. (b) Comparison of the trend of different values of $f(\alpha)$ versus $\alpha$ between among the same $\eta$, $\phi$ spaces and same trend calculated for their cross-correlation, for the $7$ TeV dataset.}
\label{alpha78e}
\end{figure*}
\textbf{Both the Figures show that -}
\begin{itemize}
\item For both the energy ranges, width of the cross-correlation curve is the maximum, followed by the width of the multifractal spectrum of the $\phi$ space and then that of the $\eta$ space.
\item Again, similar trend is observed for all the pairs of datasets in this experiment. The more wide the spectrum is the more degree of multifractality is inherent in the data series.
\end{itemize}

\begin{figure*}[h]
\centerline{
\includegraphics[width=3.5in]{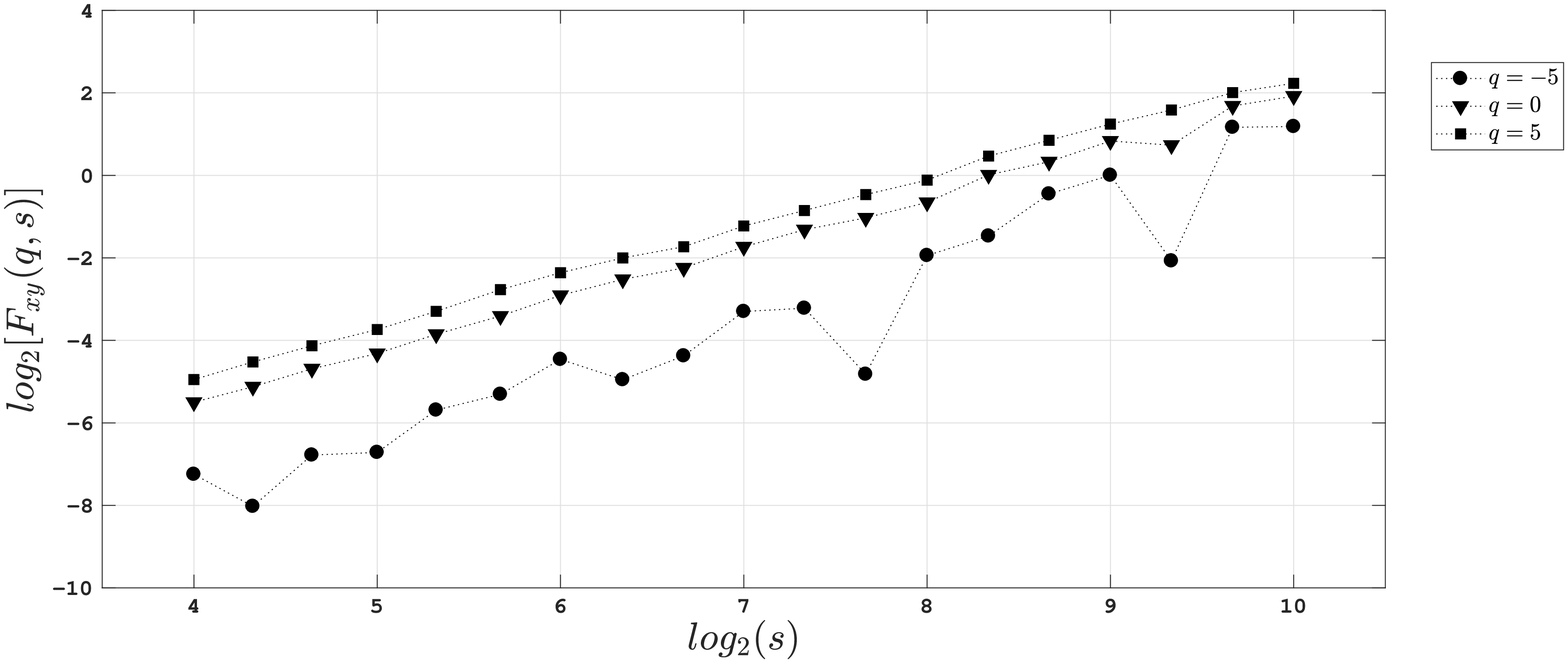}
\includegraphics[width=3.5in]{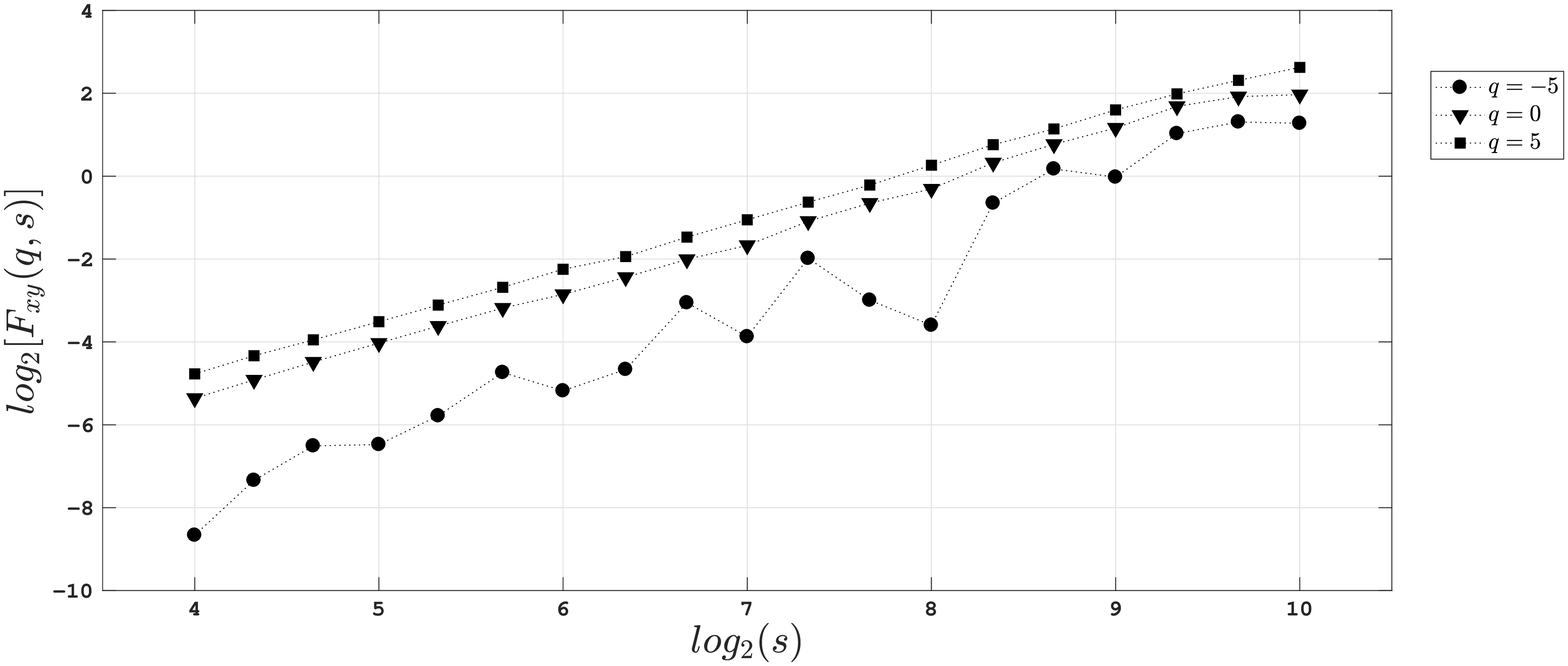}
}
\centerline{(a) \hspace*{6cm} (b)}
\caption{(a) Trend of $F_{xy}(q,s)$ vs $s$ for $q = -5,0,5$, calculated for a particular range of $\eta$ values and their corresponding $\phi$ values for $8$ TeV dataset. (b) Trend of $F_{xy}(q,s)$ vs $s$ for $q = -5,0,5$, calculated for a particular range of $\eta$ values and their corresponding $\phi$ values for $7$ TeV dataset.}
\label{mdx78efqs}
\end{figure*}

\item The $q^{th}$ order detrended co-variance $F_{xy}(q,s)$ is calculated for a particular range of $\eta$ values and their corresponding $\phi$ values as per the Step~\ref{fxy} of the MF-DXA methodology described in Section~\ref{mfdxa} and the trend of $F_q(s)$ vs $s$ for $q = -5,0,5$ is shown in the Figure~\ref{mdx78efqs}-(a) and~\ref{mdx78efqs}-(b) for $8$ and $7$ TeV datasets respectively.
\begin{itemize}
\item Their linear trend(more prominent for the values of $q>0$) confirms the power-law behavior of $F_{xy}(q,s)$ versus $s$ for all the values of $q$.
\item Similar calculation is done for all the $\eta$ ranges and their corresponding $\phi$ spaces, for both $8$ and $7$ TeV datasets and similar trend is observed.
\end{itemize}

\begin{figure*}[h]
   \begin{minipage}{0.5\textwidth}
     \includegraphics[width=3.5in]{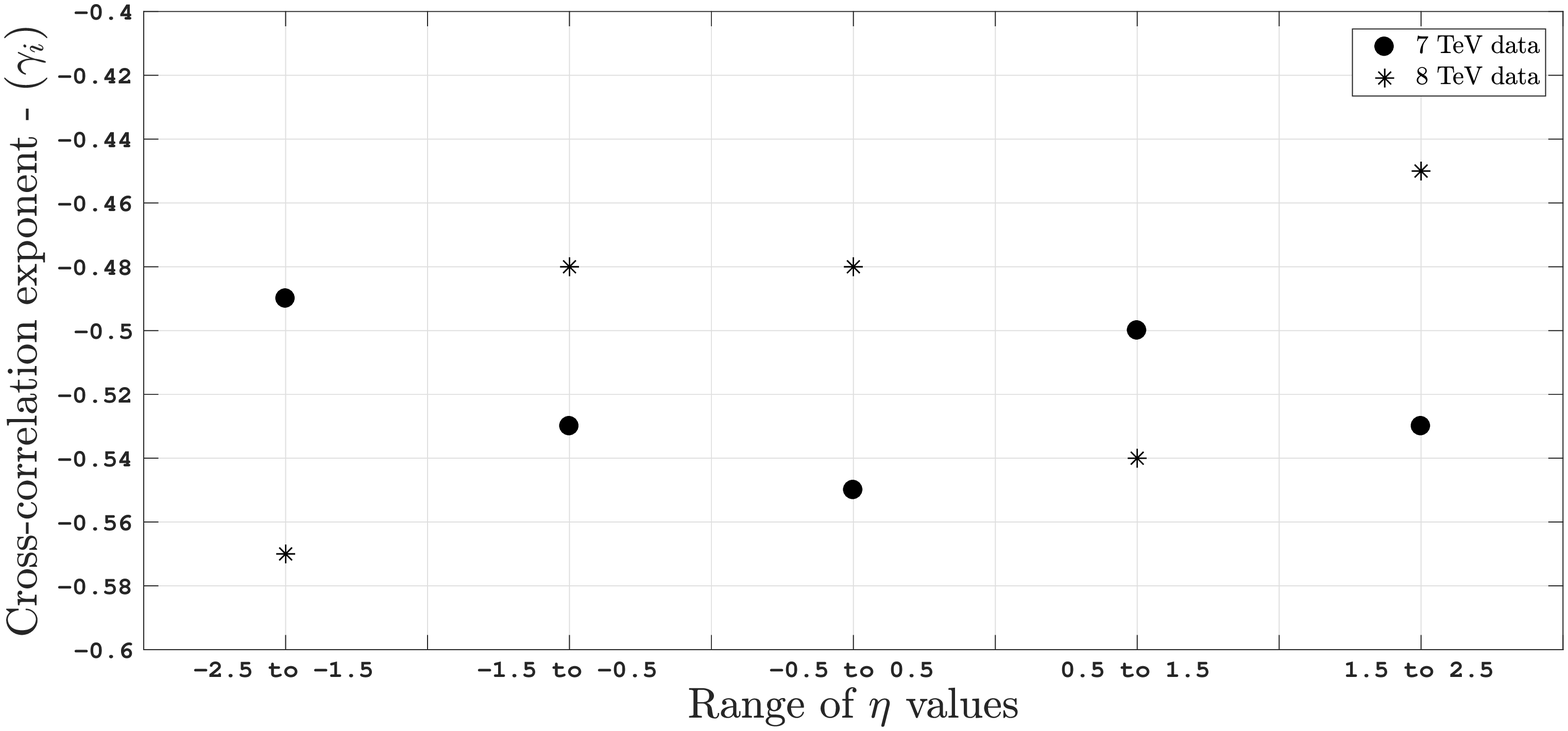}
     \caption{Comparison of Multifractal Cross-Correlation coefficient($\gamma_i$) between $\phi$ and $\eta$ spaces for all the $5$ ranges of $\eta$ values for $7$ and $8$ TeV datasets.}\label{mfdxa78}
   \end{minipage}
   \hspace{0.5cm}
   \begin{minipage}{0.5\textwidth}
   	\captionsetup{type=table}
     \begin{tabular}{@{}|l|c|c|c|c|@{}} \hline
		$\eta$ ranges&\multicolumn{4}{|c|}{\textbf{MFDXA coefficients($\gamma_i$)}}\\
		\cline{2-5}
		&\multicolumn{2}{|c|}{\textbf{$8$ TeV}}&\multicolumn{2}{|c|}{\textbf{$7$ TeV}}\\
		\cline{2-5}
		&\textbf{Orig}&\textbf{Rand}&\textbf{Orig}&\textbf{Rand}\\ 
		\hline
		$-2.5$ to $-1.5$&$-0.57$&$1.01$&$-0.49$&$0.94$\\
		\hline
		$-1.5$ to $-0.5$&$-0.48$&$1.00$&$-0.53$&$0.99$\\
		\hline
		$-0.5$ to $0.5$&$-0.48$&$1.02$&$-0.55$&$1.00$\\
		\hline
		$0.5$ to $1.5$&$-0.54$&$0.98$&$-0.50$&$1.01$\\
		\hline
		$1.5$ to $2.5$&$-0.45$&$0.96$&$-0.53$&$0.94$\\
		\hline
	\end{tabular}
     \caption{Comparison of the experimental values of Multifractal Cross-Correlation coefficients($\gamma_i$)between $\phi$ and $\eta$ spaces for all the $5$ ranges of $\eta$ values for $7$ and $8$ TeV datasets, between the original and the randomized version.}\label{gamma_mfdxa}
   \end{minipage}
\end{figure*}

\item Two sets of Multifractal Cross-Correlation coefficient, denoted by $\gamma_i$ for $i = 1,2,\ldots 5$ for each of the $8$ and $7$ TeV datasets, are computed as per the method described in Section~\ref{mfdxa}. This way, the degree of cross-correlation between $\phi$ and $\eta$ spaces for all the $5$ ranges of $\eta$ values as specified in Step~\ref{eta} for both the $8$ and $7$ TeV datasets are calculated. 

\item Then each of the azimuthal-$\phi$ spaces extracted in the Step~\ref{phi} is randomized and the Multifractal Cross-Correlation coefficients between the randomized $\phi$ spaces and the corresponding $\eta$ spaces are extracted for both the $8$ and $7$ TeV datasets, are calculated as per the method described in the Section~\ref{mfdxa}.

\item The Figure~\ref{mfdxa78} shows the comparison of Multifractal Cross-Correlation coefficients($\gamma_i$) between $\phi$ and $\eta$ spaces for all the $5$ ranges of $\eta$ values for $7$ and $8$ TeV datasets with respect to their rapidity as well as energy dependence. Here we notice that
\begin{itemize}
\item For both the $7$ and $8$ TeV data, all the $5$ $\eta$-spaces are highly cross-correlated with their corresponding $\phi$-spaces.
\item For $8$ TeV data, the first range of $\eta$-values is most cross-correlated with the corresponding $\phi$-space and the most cross-correlated range for $7$ TeV data, is the third one.
\end{itemize}

\item The comparison of $\gamma_i$-s for all the $5$ $\eta$-ranges for $7$ and $8$ TeV datasets between the original and the randomized version is shown in the Table~\ref{gamma_mfdxa}. It should be noted that $\gamma_i=1$ for uncorrelated data series. The more correlated the data series are the lower the value of $\gamma_i$. 
\begin{itemize}
\item Here the values of $\gamma_i$-s calculated for the original the randomized version differ substantially, clearly establishing the statistical significance of the results obtained from the actual data.
\end{itemize}
 
\end{enumerate}

\section{Conclusion}
\label{con}
In this work we used two rigorous and robust methodologies, namely, MF-DFA, MF-DXA analysis for the study of scaling analysis of the dynamics of the di-muon production process using dimuon data taken out from the primary dataset of RunA(2011) and RunB(2012) of the $pp$ collision at $7$TeV and $8$TeV respectively from CMS collaboration~\cite{cms2017}. We have analyzed how this scaling pattern has evolved from one rapidity range to the next one and how this change evolved from lower energy range of $7$ TeV to the higher one $8$ TeV and the findings are listed below.

\begin{enumerate}
\item The linear trend of $F_q(s)$ vs $s$ for all the values of $q$ for all the $5$ ranges of $\eta$ values for $8$ and $7$ TeV datasets confirm the fractality as well as the multifractality of all the pseudorapidity spaces. The Figures~\ref{eta78efq}-(a) and~\ref{eta78efq}-(b) show the similar trend for a particular range of $\eta$ values for both the energy ranges. Similar linear trend is observed for the $\phi$ spaces corresponding to the $\eta$ spaces, which again confirm the fractality and the the multifractality of the $\phi$ spaces as well. Figures~\ref{phi78efqs_usort}-(a) and~\ref{phi78efqs_usort}-(b) show the linear trend calculated for the $\phi$ spaces corresponding to the particular $\eta$ range for both the energy ranges.

\item The Table~\ref{gamma_mfdfa} and the Figure~\ref{mfdfa78} show how the widths of the multifractal spectrum depends from one $\eta$ space to the other and how they in turn depend from one energy range to another. It's interesting to note that for both $7$ and $8$ TeV energy the $\eta$ space corresponding to the first range of $\eta$ has the maximum width of multifractal spectrum/degree of complexity or in other words they are most multifractal in nature among the other five ranges. Moreover they have exactly the same value for the parameter. As for the minimum width of multifractal spectrum the second $\eta$ range for $8$ TeV data and third $\eta$ range for $7$ TeV data is $0.02$ which is again same for both the energy ranges.

\item The linear trend of $F_{xy}(q,s)$ vs $s$ for $q = -5,0,5$ is shown in the Figures~\ref{mdx78efqs}-(a) and~\ref{mdx78efqs}-(b) for $8$ and $7$ TeV datasets respectively, for the same specific $\eta$ range and its corresponding $\phi$ space, confirm the self-similar cross-correlation between the spaces. Similar trend is observed for rest of the $\eta$ ranges.

\item The Table~\ref{gamma_mfdxa} and the Figure~\ref{mfdxa78} show that for both the $7$ and $8$ TeV data, all the $5$ $\eta$-spaces are highly cross-correlated with their corresponding $\phi$-spaces and how the degree of cross-correlation changes from one $\eta$ space to the other and from one energy range to another. It should be noted that the degree of Multifractal Cross-Correlation - $\gamma_i$ is maximum for the first $\eta$ range for $8$ TeV data and the same is maximum for third range of $7$ TeV data. $\gamma_i$ is minimum for the fifth $\eta$ range for $8$ TeV data and for first range of $7$ TeV data.
\end{enumerate}

This analysis manifests different degree of symmetry scaling or scale-freeness in different pseudorapidity domain and at the same time different degree of cross correlation between pseudorapidity and azimuthal space at both energy. In view of this we can conclude that this new approach has the potential to provide clue to possible different dynamics behind di-muon production in different rapidity domain which may change with higher and higher energy.

\section{Acknowledgments} 
\label{ack}
We thank the \textbf{Department of Higher Education, Govt. of West Bengal, India} for logistics support of computational analysis.


\end{document}